\documentclass[%
 aip,
 jmp,%
 amsmath,amssymb,
preprint,%
]{revtex4}

\pdfoutput=1

\usepackage{amssymb}
\usepackage{todonotes}

\usepackage{lineno}

\usepackage{algpseudocode}
\usepackage{algorithm}
\usepackage{listings}
\usepackage{graphicx}
\usepackage{subfig}

\begin{document}

\title{readPTU: a Python Library to Analyse Time Tagged Time Resolved Data}

\author{G. C. Ballesteros}
\email{gb173@hw.ac.uk}
\author{R. Proux}%
\author{C. Bonator}%
\email{c.bonator@hw.ac.uk}
\author{B. D. Gerardot}%
\email{b.d.gerardot@hw.ac.uk}
\affiliation{Institute of Photonics and Quantum Sciences (IPaQS), Heriot-Watt University, Edinburgh, EH14 4AS, UK}
 
\homepage{http://qpl.eps.hw.ac.uk/}

\date{\today}

\begin{abstract}
readPTU is a python package designed to analyze time-correlated
single-photon counting data. The use of the library promotes the storage
of the complete time arrival information of the photons and full
flexibility in post-processing data for analysis. The library supports the
computation of time resolved signal with external triggers and second
order autocorrelation function analysis can be performed using multiple
algorithms that provide the user with different trade-offs with regards to
speed and accuracy. Additionally, a thresholding algorithm to perform time
post-selection is also available.  The library has been designed with
performance and extensibility in mind to allow future users to implement
support for additional file extensions and algorithms without having to
deal with low level details. We demonstrate the performance of readPTU by
analyzing the second-order autocorrelation function of the resonance
fluorescence from a single quantum dot in a two-dimensional semiconductor.
\end{abstract}

\maketitle

\section{Motivation and significance}

Time-correlated single-photon counting (TCSPC) experiments have found
widespread applications in different disciplines, including  the
characterization of individual optical emitters~\cite{lounis2000single,
PhysRevLett.116.020401}, advanced microscopy ~\cite{niehorster2016multi,
pian2017compressive}, Bell's inequalities verification using astronomical
sources ~\cite{PhysRevLett.118.060401}, or the real time tracking of
physiochemical reactions~\cite{suhling2005time}. This technique is also
a major tool for the characterization of single photon emission
~\cite{PhysRevLett.116.020401, dada2016indistinguishable,
PhysRevLett.113.113602}, relevant for quantum technologies such as linear
optics quantum computing \cite{ladd2010quantum} and spin-photon interfaces
for quantum communication networks \cite{gao2012observation,
togan_quantum_2010}. TCSPC measurements can be used to characterize the
statistical properties of the emitted light \cite{fox2006quantum} and it is
thus a tool to prove the single-photon nature of a source and provide
useful information about its internal dynamics.


When recording TCSPC data, one usually only directly computes the quantities of
interest, such as a histogram of the delays between photon clicks at two
channels, since storing the timing information for all detector clicks can
easily grow above gigabyte sizes. On the other hand, storing this
information can be advantageous to perform more sophisticated data
analysis. For example, time traces of the detected detector click rates
can reveal intensity spikes and the emitters' dynamics, or enable post
selection of specific time intervals. The readPTU\footnote{The repository
for the readPTU is hosted at:
\emph{https://github.com/qpl-public/readPTU}} library presented here
enables the researcher to achieve this, providing added flexibility in
their experiments.





The library has been written as a Python module that interfaces with
a C library that handles the most computing-intensive aspects. This
combination provides a user-friendly interface to a high performance set
of underlying routines. The underlying C library provides functionalities
to efficiently read a stream of detector click records, with no required
knowledge about the low level details of how the information is encoded on
the binary files, making the library of algorithms easy to extend.


Our hope is that the work presented here will allow researchers across a
variety of disciplines to analyze raw TCSPC data more efficiently.

\section{Autocorrelation measurements using TCSPC}
\label{section:theory}
\begin{figure*}
  \centering
    \includegraphics[width=0.8\textwidth]{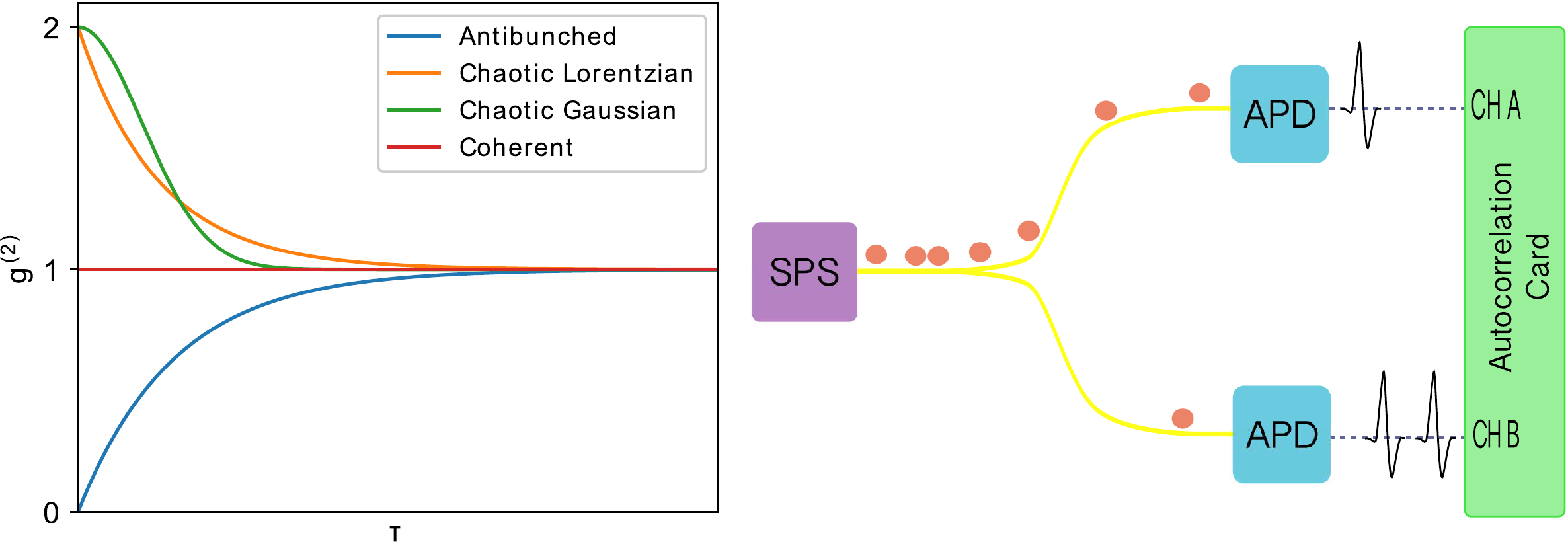}
    \caption{(Left) Typical second order autocorrelation functions for
    different types of sources. Only antibunched light will show a dip
going to zero at $\tau=0$ indicating that two photons never arrived
simulatenausly at both channels. (Right) Experimental setup to measure the
second order autocorrelation from a single photon source.}
\label{fig:typical_g2s} \end{figure*}

One of the main applications of TCSPC is the characterization of the
statistical properties of a light source, such as the anti-bunched nature
of single photon sources, through the second order autocorrelation of the
field. The right panel of Fig.~\ref{fig:typical_g2s} shows a simplified
setup of such an experiment. The emitted photon field is split by a 50/50
beam splitter and directed to two single photon detectors, such as
avalanche photo diodes (APS) or super conducting nanowire single photon
detectors (SNSPD). Upon photon detection, detector clicks are recorded by
channels A and B of time-Correlated single photon counting system
(TCSPC). Since a single photon source cannot produce a click on both
channels simultaneously, a dip is observed in the coincidence rate when
the delay between the channels is equal (Fig.~\ref{fig:typical_g2s}, left
panel). This result is only possible under the assumption that the
electromagnetic field is quantized \cite{fox2006quantum}. These results
can be extended to the n-th order correlation functions and provide a full
characterization of the coherence properties of an electromagnetic field. 

In this section we will introduce the basic theory behind the second order
autocorrelation function and how it can be experimentally measured using TCSPC.
To compute the second order autocorrelation function, $g^{(2)}(\tau)$, we need
to consider first what is the probability amplitude for measuring a photon with
a delay of $\Delta t$ in detector B after a having measured one in detector
$A$. This is given by~\cite{glauber1963coherence}:

\begin{equation}
\langle f| E_A^{(+)}(t)E_B^{(+)}(t+\Delta t)|i \rangle
\end{equation}

where $E^{(+)}(r, t)$ is the photon annihilation operator and $|i \rangle$ and
$|f \rangle$ are the initial and final states respectively. Similarly we can
define the probability of detecting two photons in different detectors
$(r=A,B)$. We will show in Sect.~\ref{sec:derivation} how considering all final
states and averaging over the ensemble of initial states leads to the definition
of the second order autocorrelation function $G^{(2)}(\tau)$: 

\begin{equation}
    G^{(2)}(\tau) = \langle E^{(-)}(0)E^{(-)}(\tau)E^{(+)}(\tau)E^{(+)}(0) \rangle
\end{equation}

In the latter expressions we have fixed $t=0$ under the assumption that we are
dealing with a stationary process, in that the value of the autocorrelation
function will only depend on time differences.






\subsection{From photon statistics to $g^{(2)}(\tau)$}
\label{sec:derivation}
We show now how building a histogram of time delays between photons in
channel A and channel B of the TCSPC system gives us access to the second
order autocorrelation function of a stream of photons. Put another way,
this histogram gives an approximation to the joint probability
distribution of measuring a photon in the first detector in between times
$t + \mathrm{d}t$ and $t + \tau + \mathrm{d}t$ on the second
detector~\cite{proux:tel-01278553}. The probability per unit time to
obtain the final state $|f\rangle$ after the detection of a photon at $t$
in channel A and at $t+\tau$ in channel B from the initial state
$|i\rangle$ is proportional to:

\begin{equation}
    w_{2,if}(\tau) = |\langle f | E ^{(+)}(\tau) E^{(+)} (0) | i \rangle|^2
    \label{eqn:g2_def}
\end{equation}

If we sum over all possible final states and introduce the density operator
($\rho=\sum_{i'} P_{i'}|i'\rangle \langle i'|$) to consider the mixture of
initial states we conclude that:

\begin{equation}
    w_2(\tau) = Tr[\rho E^{(-)} (0) E ^{(-)}(\tau) E^{(+)}(\tau) E^{(+)}(0)] = G^{(2)}(\tau)
\end{equation}

\subsection{Algorithms}
readPTU provides multiple algorithms to postprocess TCSPC data, each featuring
different trade-offs.

\subsection{Intensity time trace}

\begin{algorithm}
\label{alg:timetrace}
\caption{Timetrace Algorithm}
\begin{algorithmic}
\Procedure{TimeTrace}{}
    \State $\mathrm{EndOfBin} \gets \texttt{dT}$
    \State $n \gets 0$
    \State $\mathrm{CurrBin} \gets 0$
    \While{$n < \mathrm{NumRecrods}$}
        \State $(\mathrm{channel}, t) \gets \mathrm{ParseNextRecord}$

        \If {$t < \mathrm{EndOfBin}$}
            \State $n \gets n + 1$
        \ElsIf {$t > \mathrm{EndOfBin}$}
            \State $\mathrm{UpdateTimeTrace}(n, \mathrm{CurrBin})$
            \State $\mathrm{EndOfBin} \gets \mathrm{EndOfBin} + \texttt{dT}$
            \State $n \gets 1$
            \State $\mathrm{CurrBin} \gets \mathrm{CurrBin} + 1$
        \EndIf
    \EndWhile
\EndProcedure
\end{algorithmic}
\end{algorithm}

The first algorithm provided by readPTU enables user to obtain
a time-trace of the photon count. Its pseudo-code is shown in Alg.~1. The
algorithm goes through the recorded photon events and assigns them to the
current time bin as long as its time-tag falls within it. Once the time
tag is larger than the end time of the current time bin, CurrBin is
updated and the next time bin starts to be filled. The length of each time
bin sets the trade-off between time resolution and the error in the photon
count due to Poissonian statistics. The intensity time-trace can be used
to study emitter dynamics as will be demonstrated in
Sect.~\ref{sec:example}.

\subsection{$g^{(2)}(t)$ Algorithms}
\subsubsection{Naive Algorithm}
\label{sec:naive_algo}
\begin{figure*}
  \centering
    \includegraphics[width=0.8\textwidth]{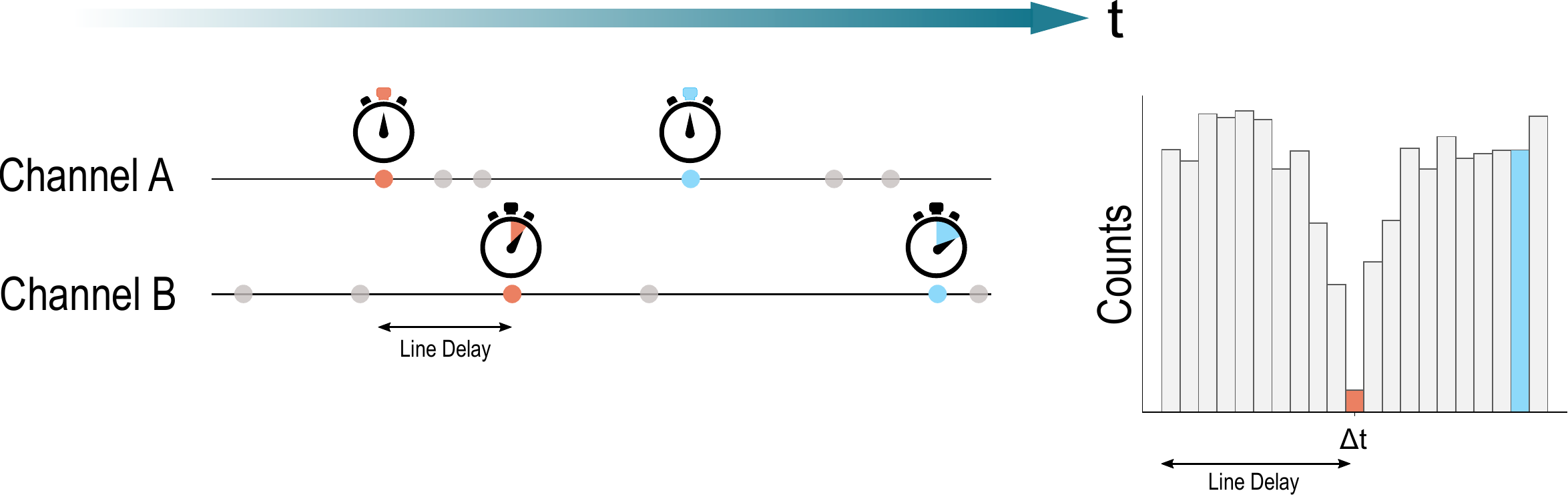}
    \caption{Schematic representation of the naive algorithm. When the first
    (red) photon arrives at channel A, a stopwatch is started. When a photon
    arrives at channel B, the stopwatch is stopped and the delta introduced in
    a histogram. After the stop photon in channel B has arrived we wait for
    another photon to arrive at channel A and repeat the process. Notice how
    all the greyed out photons haven't been used to compute the autocorrelation
    function. This leads to a less efficient use of the data available and an
    exponential decaying artifact on the computed solution.}

\label{fig:naive_algo} \end{figure*}

The derivation in section~\ref{sec:derivation} shows how $g^{(2)}(\tau)$ can be
approximated by a histogram of the delays between photon clicks in channels A
and B of the autocorrelation card. The simplest implementation of this idea is
described in Fig.~\ref{fig:naive_algo} and  Algorithm~2.

When a photon is detected in channel A, a stopwatch is started. When, after a
time delay $\mathrm{d}t$ a second photon is recorded at channel B,
$\mathrm{d}t$ is assigned to the histogram in the call to
\textit{UpdateHistogram} and the process repeated. The \textit{UpdateHistogram}
function additionally checks if the measured delay is within a user defined
interval (the autocorrelation window) that establishes the longest
delay that will be stored in the output histogram.

An artificial delay between the two channels is typically added to shift $t=0$
from the origin, so that $g^{(2)}(t=0)$ can be see when running the
autocorrelator in histogram mode. Reversing start and stop channels brings the
$t=0$ autocorrelation out of the histogram: readPTU library has an operation
mode that removes this issue by running an algorithm symmetric on the
assignment of start and stop channels. Although this algorithm is extremely
fast and good enough for quickly exploring the data it has two major
shortcomings. First, the time delay between two consecutive photons follows an
exponential distribution. Since the algorithm described above only looks at
consecutive pairs, an exponential decay artifact is introduced. Second, photons
in channel A are ignored while waiting for a click in channel B, wasting
available information (see the photons greyed out in
Fig.~\ref{fig:naive_algo}). 

\begin{algorithm}
\label{alg:naive}
\caption{Naive Algorithm}
\begin{algorithmic}
\Procedure{Naive}{}
    \State $\mathrm{WaitForStop} \gets \texttt{FALSE}$
    \State $n \gets 0$
    \While{$n < \mathrm{NumRecrods}$}
        \State $(\mathrm{channel}, t) \gets \mathrm{ParseNextRecord}$
        \State $n \gets n + 1$

        \If {$\mathrm{channel}==\mathrm{start} \; \mathrm{AND} \; \mathrm{NOT} \; \mathrm{WaitForStop}$}
            \State $\mathrm{StartTime} \gets t$
            \State $\mathrm{WaitForStop} \gets \texttt{TRUE}$
        \ElsIf {$\mathrm{channel}==\mathrm{stop} \; \mathrm{AND} \; \mathrm{WaitForStop}$}
            \State $\mathrm{d}t \gets t - \mathrm{StartTime}$
            \State $\mathrm{UpdateHistogram}(\mathrm{d}t)$
            \State $\mathrm{WaitForStop} \gets \texttt{FALSE}$
        \EndIf
    \EndWhile
\EndProcedure
\end{algorithmic}
\end{algorithm}

\subsubsection{Ring Algorithm}
The ring algorithm presented here (pseudo-code in Algorithm~3)
fixes both shortcomings highlighted in the previous section in a numerically
efficient way. Click times for channel A are stored in a buffer. Whenever
detector B clicks, the algorithm loops over the click times stored in the
buffer and generates a list of time delays, which are then stored in the
histogram. Each time a photon clikcs at channel A, if the buffer is full, the
oldest photon click is replaced for the newly detected, hence the ring (buffer)
name. Even though this greatly mitigates the exponential artifacts shown by the
naive algorithm, a situation may occur where a time delay is larger than the
correlation window. This situation is corrected by keeping track of the longest
seen delay on a loop over the start photon buffer. If a delay bigger than the
correlation window is encountered the buffer is doubled in size.


\begin{algorithm}
\label{alg:ring}
\caption{Ring Algorithm}
\begin{algorithmic}
\Procedure{Ring}{}
    \State $\mathrm{WaitForStop} \gets \texttt{FALSE}$
    \State $n \gets 0$
    \While{$n < \mathrm{NumRecrods}$}
        \State $(\mathrm{channel}, t) \gets \mathrm{ParseNextRecord}$
        \State $n \gets n + 1$

        \If {$\mathrm{channel}==\mathrm{start}$}
            \State $\mathrm{StorePhoton}(t)$
        \ElsIf {$\mathrm{channel}==\mathrm{stop}$}
            \For {$t_\mathrm{start}$ in PhotonBuffer}
                \State $\mathrm{d}t \gets t - t_\mathrm{start}$
                \State $\mathrm{UpdateHistogram}(\mathrm{d}t)$
            \EndFor
        \EndIf

    \EndWhile
\EndProcedure
\end{algorithmic}
\end{algorithm}



\subsection{readPTU}
readPTU is distributed as a Python library with a compiled C component,
running under Windows, MacOS and Linux. Its main goal is to provide
a user-friendly interface via Python to users wanting to postprocess
multi-gigabyte files of TCSPC data. It currently supports calculation of
intensity time traces and  $g^{(2)}$ autocorrelations (with/without
post-selection).

The main C file acts as a template for (currently) 3 dynamically linked
libraries. Each of the libraries provides support for a different input
file format. Currently T2 mode files for the HydrapHarp, TimeHarp,
PicoHarp and HydraHarp2 devices from PicoQuant are supported.
Instructions on how to add new files formats is explained in the package
documentation. Each file format requires a parser to be defined in
\textit{parsers.c}. Through using different libraries, the overhead from
dynamically selecting (via function pointers or switch statements) what
specific file format parser to use is completely removed, while keeping
a single codebase and an easily extensible system for future file formats.
The whole library has been designed to primarily target performance: to
this end, buffered input and output is used by reading multiple records
from the TCSPC files simultaneously. This optimization reduces by three
orders of magnitude the number of file read operations required to process
the input. Memory locality was taken into account by substituting the use
of linked lists for contiguous memory arrays. Additionally, all the
functions provided by the library offer multithread support to take
advantage of modern multi-core CPUs.  The parallelization strategy is
based on analyzing different ranges of records in the TCSPC files
separately and the combining the results together. This means that there
are no shared variables among the different threads and the algorithms
don't require the use of synchronization primitives. Detector click times
are time-ordered in the buffers used by Alg.~3; this allows one to
introduce a check to break out of the histogram updating loop as soon as
a time delay longer than the one that can fit in the histogram is
detected.

\subsubsection{Sample usage}
readPTU is designed to be as user-friendly as possible. The code listing below
presents an example of how to use the library to obtain a timetrace and compute
the $g^{(2)}(\tau)$ from TCSPC data.

Files are opened within \textbf{With} block to make sure that the file is
appropriately closed when no longer needed. The \textit{timetrace} function has
only one mandatory argument, the size of the time bins over which the clicks
are averaged to compute the intensity trace. Smaller bins provide a greater
time resolution at the expense of less accuracy due to the stochastic nature of
the emission and detection process. Additionally, one can specify which
autocorrelator channel should be used to build the time trace if the user wants
to separate the counts from the different channels. Options are available to
limit the range of records. Finally, the computation of the time trace can be
parallelized by specifying the number of threads to be used.



\begin{figure*}[!ht]
     \subfloat[]{%
       \includegraphics[width=0.45\textwidth]{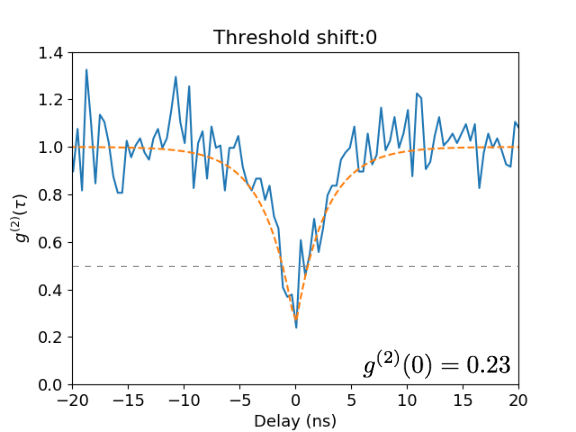}
     }
     \hfill
     \subfloat{%
       \includegraphics[width=0.45\textwidth]{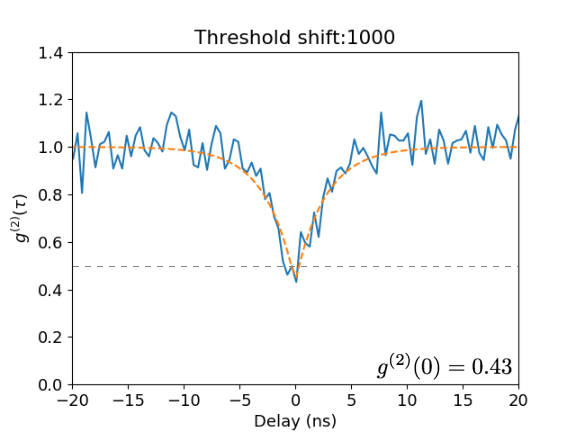}
     }
     \\
     \subfloat{%
       \includegraphics[width=0.45\textwidth]{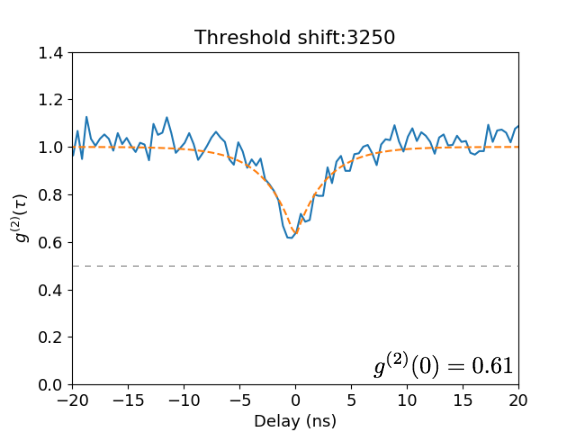}
     }
     \hfill
     \subfloat{%
       \includegraphics[width=0.45\textwidth]{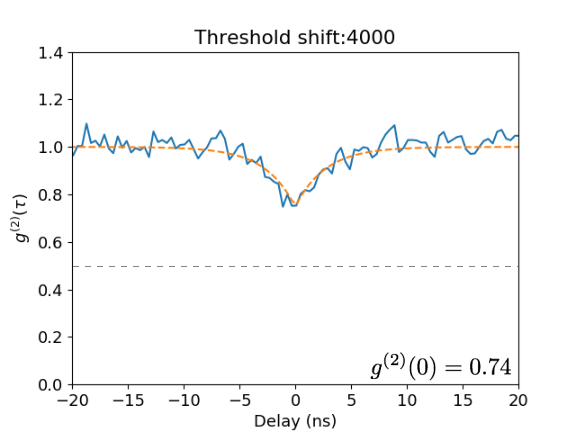}
     }

     \caption{Second order autocorrelation as a function of the applied
     threshold. The legend on each plot shows the minimum $g^{(2)}(0)$
     reached. The figure titles indicates the downwards shift in Hz that
     was applied to the optimum threshold function. As the threshold is
     lower more background photons are allowed into the second order
     autocorrelation computation decreasing the $g^{(2)}(0)$ value. When
     the threshold is shifted downwards by 4000 counts or more we obtain
     the same result regardless of the amount of postselection performed.}

  \label{fig:g2_fit}
   \end{figure*}

The computation of the second order autocorrelation is available via the
\textit{calculate\_g2} function. In this case, the only mandatory argument is
the length of the correlation window. The optional
arguments allow the user to specify the time resolution of the histogram. The
default value in this case is to split the histogram into 1024 bins if a time
in seconds is not specified. The argument \textit{post\_selec\_ranges} takes a
list of pairs of values with start and stop records. Only clicks belonging to
records within one of those ranges are used to compute $g^{(2)}$. As discussed
above, this option allows users, for example, to exclude time ranges where they
suspect photons likely only correspond to laser background. readPTU provides
the function \textit{construct\_postselect\_vector} to automatically generate
ranges based on a hard threshold. The script used to analyze and produce the
plots in Fig.~\ref{fig:g2_fit} is available on the code repository together with
instructions on how to extend the library to manipulate other file formats.

\section{Example: resonance fluorescence of a quantum emitter with spectral fluctuations}
\label{sec:example}
One application of the functionalities offered by readPTU is to study the
dynamics  of single photon emitters. For solid-state emitters, fluctuations in
the surrounding environment result in randoms shifts of the absorption/emission
wavelength~\cite{PhysRevLett.108.107401}. When exciting narrow absorption lines
with a narrowband lasers, these random shift bring the emitter in and out of
resonance with the laser, resulting in intermittent optical emission
\cite{santana2017generating, Kumar:16}. When the emitter is not resonant with
the laser, only background light is collected, degrading the Signal to
Background Ratio (SBR), an important parameter for antibunching
experiments. In resonance fluorescence, the background laser can be
filtered by polarization from the signal~\cite{kuhlmann2013dark}, but it
remains challenging to completely remove it. This problem can be solved by
post-selecting such that only the photons detected in time intervals when
the emitter was resonant with the laser, i.e. when the emission rate is
above a given threshold, contribute to the  $g^{(2)}(\tau)$ measurement.

Here we show how readPTU can be used to analyze TCSPC data from a single
photon emitter in an atomically-thin monolayer of WSe$_2$ \cite{Kumar:16},
at cryogenic temperature. The emitters were addressed by resonance
fluorescence in a confocal microscope. The laser linewidth is 10 kHz,
while the emitter (typical linewidth $\approx 100$~MHz) wavelength
fluctuates on 10 GHz scale. While the timescales of the spectral
fluctuations vary, they are relatively slow with dynamics longer
than a second which can be easily post-selected for.

\begin{figure*}
   \centering
     \includegraphics[width=0.8\textwidth]{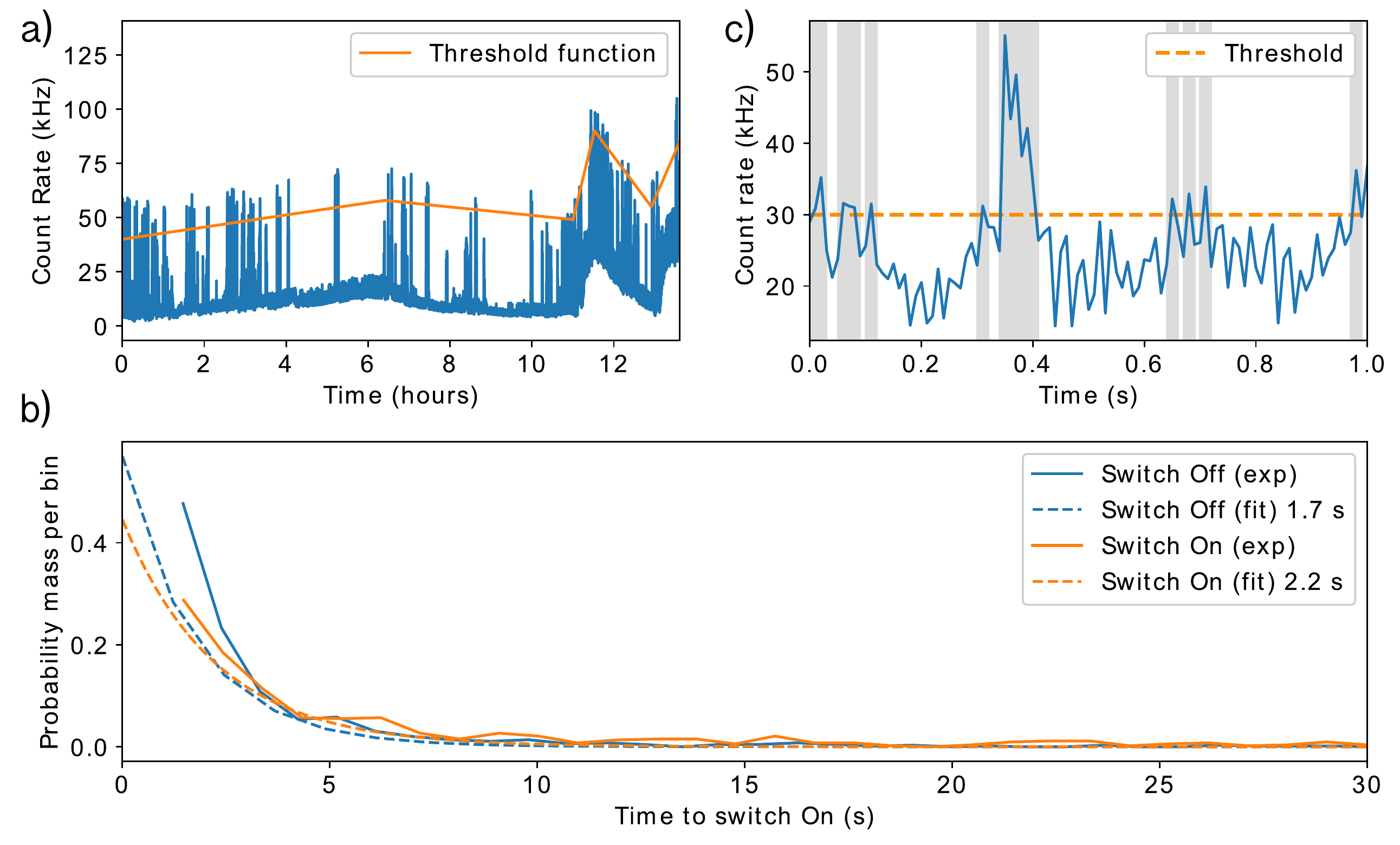}
     \caption{a) Intensity time trace of resonance fluorescence from a quantum
     dot in monolayer WSe$_2$. As the emitter wavelength jitters, it goes in
     and out of resonance randomly. When the emitter is out of resonance, only
     background laser photons are collected. By applying time post-selection, only
     photon count rates above a certain threshold are employed to compute the
     autocorrelation function. b) Zoom in of the intensity trace shown in a).
     Only photons collected during the highlighted windows of time are used to
     compute $g^{(2)}(\tau)$. c) Probability distribution for the duration of
     the time intervals during which the emitter is in the on/off state.}
     
\label{fig:threshold} \end{figure*}

First, we extract an intensity timetrace. To this end, the user only has to
specify the size of time bin over which the number of photons will be averaged,
optionally the channel number. The timetrace helps us identify problems during
the experiment, such as drifts of the experimental setup or blinking, as can be
observed in Fig.~\ref{fig:threshold}. When the emitter goes out of
resonance, the count number falls drastically to the level of the
background produced by the uncancelled excitation laser. It can also be
observed how the background laser cancellation fidelity drifts throughout
the experiment. We use this information to design a piece-wise linear
post-selection thresholding function adapted to our data.

Once a threshold has been established, we can look at the distribution of the
duration of periods for which the emitter has been in/out of resonance. This
shows that the emitter was above threshold for just 0.5\% of the duration of
the experiment, due to the presence of  charge traps that perturb the emitter's
electrical environment. The dynamics of the short lived charge trap states are
shown in Fig.~\ref{fig:threshold}. As expected  the distribution of time
intervals for which the emitter is in the on (off) state follow approximately
an exponential distribution with a mean lifetime of 1.75 s (2.24 s).

Since only laser light is detected most of the time, the mininmum achieavable
$g^{2}(0)$ is limited by the SBR averaged over the whole duration of the
experiment, which is much smaller than the peak SBR. By post-selecting with the
threshold function shown in Fig.~\ref{fig:threshold}, the minimum $g^{2}(0)$ is
improved from 0.74 to 0.23 which demonstrates that the observed system is
really a single emitter, since $g^{(2)}<0.5$ \cite{fox2006quantum}.



\section{Conclusions}
\label{sec:conclusions}
Here we present the implementation of multiple algorithms to compute the second
order autocorrelation function from TCSPC data. Lifetime measurements can also
be performed by choosing the appropiate parameters. The use of raw TCSPC data
allows us to perform time post-selection of the data based on the intensity.
This can be used to significantly improve experimental results in the resence
of emitter blinking or spectral fluctuations \cite{Kumar:16}. Thanks to the
focus on performance and extensibility multi-GB files can be analyzed in a few
seconds and new algorithms can be implemented without having to worry about low
levels details. More importantly, we have found that keeping raw TCSPC data has
helped us improve our experimental setups and data analysis routines and having
a library capable of efficiently analyzing it has promoted that it is always
stored.

\section{Funding Information}

This work has been supported by the Engineering and Physical Sciences Research Council (EPSRC) under the grants: EP/I023186/1, EP/L015110/1, EP/S000550/1 and by the European Research Counil (ERC) under grant number 725920. B.D.G. thanks the Royal Society for a Wolfson Merit Award and the Royal Academy of Engineering for a chair in Emergent Technologies.





\begin{figure*}
  \centering
    \includegraphics[width=0.8\textwidth]{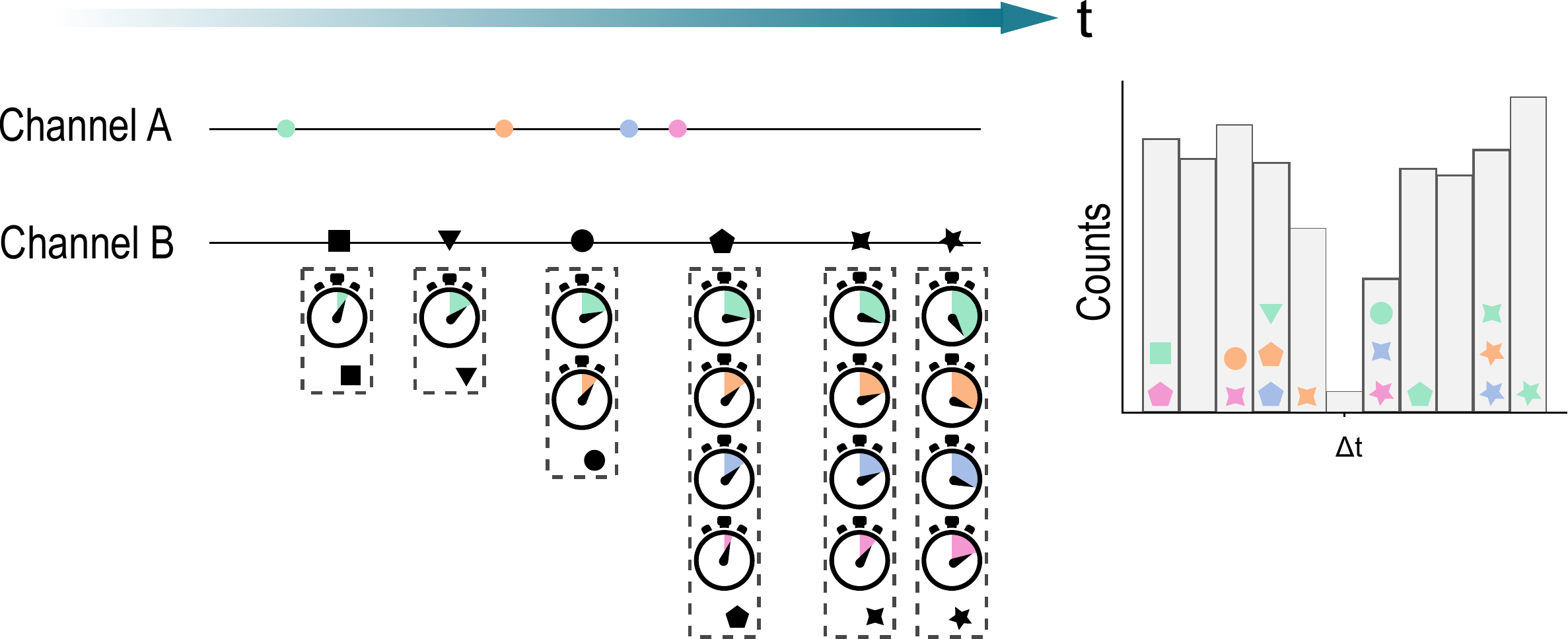}
    \caption{Schematic representation of the ring algorithm. By
    introducing a buffer for the start channel (A) we can compute delays
    with respect to all the photons arriving at channel B. Every time
    a photon is detected on channel B a set of delays with respect to all
    the photons in channel's A buffer is computed. The computed delays are
    represented in the boxes below each photon in Channel B. With the
    naive algorithm it would not have been possible to use the triangle
    photon (ch B) as we would have been waiting idly to detect a photon on
    channel A. It also would not have been possible to use the pink photon
    (ch A) as we would have been waiting for a stop photon in channel B.}

\end{figure*}

\end{document}